\documentclass[a4paper,11pt]{article}

\usepackage{jinstpub} 

\usepackage[normalem]{ulem}
\usepackage{mathcomp}
\usepackage{upgreek}
\usepackage[group-minimum-digits=4, group-separator={,}]{siunitx}
\newcommand{\bm}[1]{\boldsymbol{#1}} 
\renewcommand{\mathord}{}

\title{\boldmath Laser Thomson Scattering Measurements around Magnetized Model in Rarefied Argon Arcjet Plume}

\author[a,1]{H. Katsurayama,\note{Corresponding author.}}
\author[a]{R. Wada,}
\author[a]{K. Moriyama,}
\author[b]{and K. Tomita}

\affiliation[a]{Mechanical and Aerospace Engineering Course, Graduate School of Sustainability Science, Tottori University,\\4-101 Koyama-cho Minami, Tottori, Tottori 680-8552, Japan}
\affiliation[b]{Division of Quantum Science and Engineering, Graduate School of Engineering, Hokkaido University,\\Kita 13, Nishi 8, Kita-ku, Sapporo, Hokkaido 060-8628, Japan}

\emailAdd{katsurayama@tottori-u.ac.jp}

\abstract{To elucidate the role of the Hall effect in magnetohydrodynamic (MHD) aerobraking in rarefied flows,we measured the radial distributions of electron temperature and density in front of a magnetized model in a rarefied argon arcjet wind tunnel using the laser Thomson scattering method. We also developed a water-cooled magnetized model to prevent thermal demagnetization during the measurement. The measured electron density distributions were in excellent agreement with computational fluid dynamics (CFD) predictions. It was also found that the magnetic field had little effect on the electron density distribution around the model. In the case without the magnetic field, the measured electron temperature almost agreed with the CFD prediction. However, the measured electron temperature increase caused by applying the magnetic field was about \SI{1000}{K} less than that of the CFD prediction. This discrepancy indicates that the location of an insulating boundary in the plasma is far from the model.}


\arxivnumber{2312.09593} 

\proceeding{20$^{\text{th}}$ International Symposium on Laser-Aided Plasma Diagnostics\\
  Kyoto, Japan\\
  10-14 September 2023}

\begin{document}
\maketitle
\flushbottom

\section{Introduction}
\label{sec:intro}

\begin{figure}[b]
  \centering
  \includegraphics[height=50mm,clip]{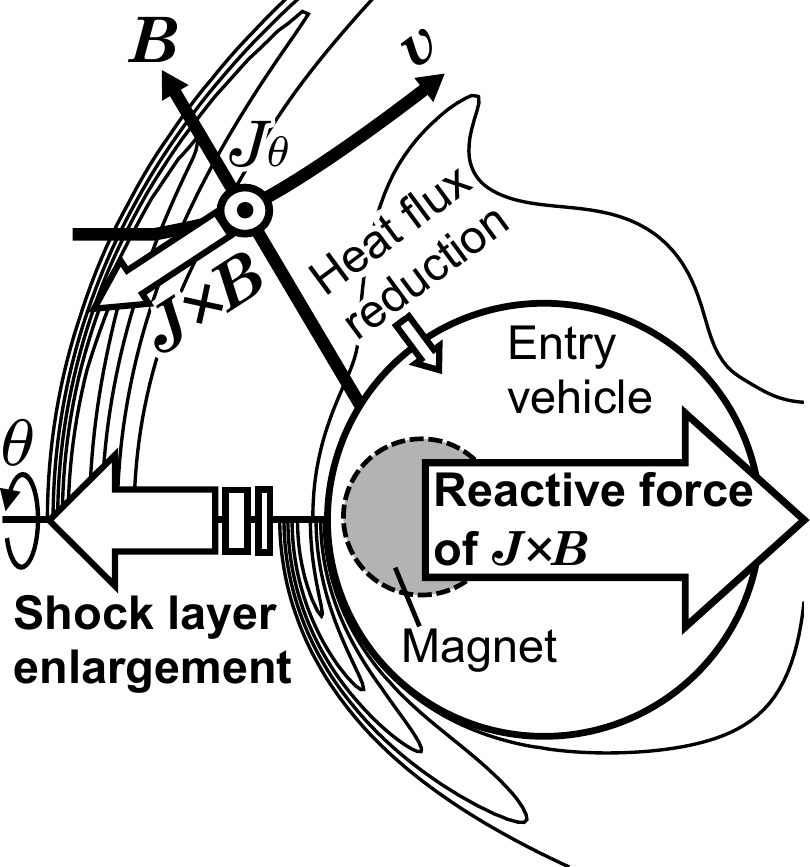}
  \caption{\label{fig:MHD} Schematic of the MHD aerobraking.}
\end{figure}

Magnetohydrodynamic (MHD) aerobraking~\cite{1,2,3,4} (figure~\ref{fig:MHD}) is expected to be a revolutionary system that can actively reduce the aerodynamic heating of atmospheric entry vehicles. This system applies a magnetic field $\bm{B}$ to the weakly ionized shock layer in front of the entry vehicle. The interaction (Ohm's law) between $\bm{B}$ and the weakly ionized flow with velocity $\bm{v}$ generates a circumferential current $J_\theta$. The further interaction between $\bm{B}$ and $J_\theta$ produces the Lorentz force $\bm{J}\mathord\times\bm{B}$ in the opposite direction of the flow. The Lorentz force enlarges the shock layer (MHD shock layer enlargement), and its reaction force decelerates the vehicle. As a result, the system can significantly reduce aerodynamic heating along the entire entry trajectory.

For this system to be effective, the magnetic interaction parameter $Q$~\cite{5}, which is the ratio of $\bm{J}\mathord\times\bm{B}$ to the inertial force $\rho_\infty V_\infty^2$, must be greater than 1:
\begin{equation}
  \label{eq:Q}
  Q = \frac{\bm{J}\times\bm{B}D}{\rho_\infty V_\infty^2} = \frac{J_\theta |\bm{B}| D}{\rho_\infty V_\infty^2}\gg 1
\end{equation}
where the vehicle diameter $D$ is the representative length of the system, and $\rho_\infty$ and $V_\infty$ are free stream density and velocity. Assuming negligible ion slip effect, circumferential velocity ($v_\infty\mathord\sim 0$), and electron pressure gradient ($\nabla p_e\mathord\sim 0$), we can define $J_\theta$ according to the generalized Ohm’s law as follows~\cite{6},
\begin{equation}
  \label{eq:J_theta}
  J_\theta \simeq \frac{\sigma}{1+\beta^2}\left[-\beta\frac{\left(\bm{E}\times\bm{B}\right)_\theta}{|\bm{B}|}+\left(\bm{v}\times\bm{B}\right)_\theta\right]
\end{equation}
where $\sigma$ is the electric conductivity, $\beta$ is the Hall parameter defined by $\beta\mathord=\sigma|\bm{B}|/e n_e$ with the electron number density $n_e$ and the elementary charge $e$, and $\bm{E}$ is the Hall electric field. In high-density regions (i.e., low altitudes) where the Hall effect is small ($\beta\mathord\sim 0$), $Q$ takes the simple form $Q\mathord=\sigma|\bm{B}|^2 D/\rho_\infty V_\infty$ because of $J_\theta\mathord\sim\sigma\left(\bm{v}\mathord\times\bm{B}\right)_\theta\mathord\sim \sigma V_\infty |\bm{B}|$. On the other hand, in low-density regions (i.e., high altitudes) where the Hall effect is large ($\beta\mathord\gg 1$), $Q$ is proportional to $\bm{E}$ because of $J_\theta\mathord\sim\mathord-\frac{\sigma}{\beta}\frac{\left(\bm{E}\mathord\times\bm{B}\right)_\theta}{|\bm{B}|}\propto\bm{E}$. Therefore, the generation of $\bm{E}$ is essential for the MHD aerobraking from a high altitude.

\begin{figure}[b]
  \centering 
  \includegraphics[height=39mm,clip]{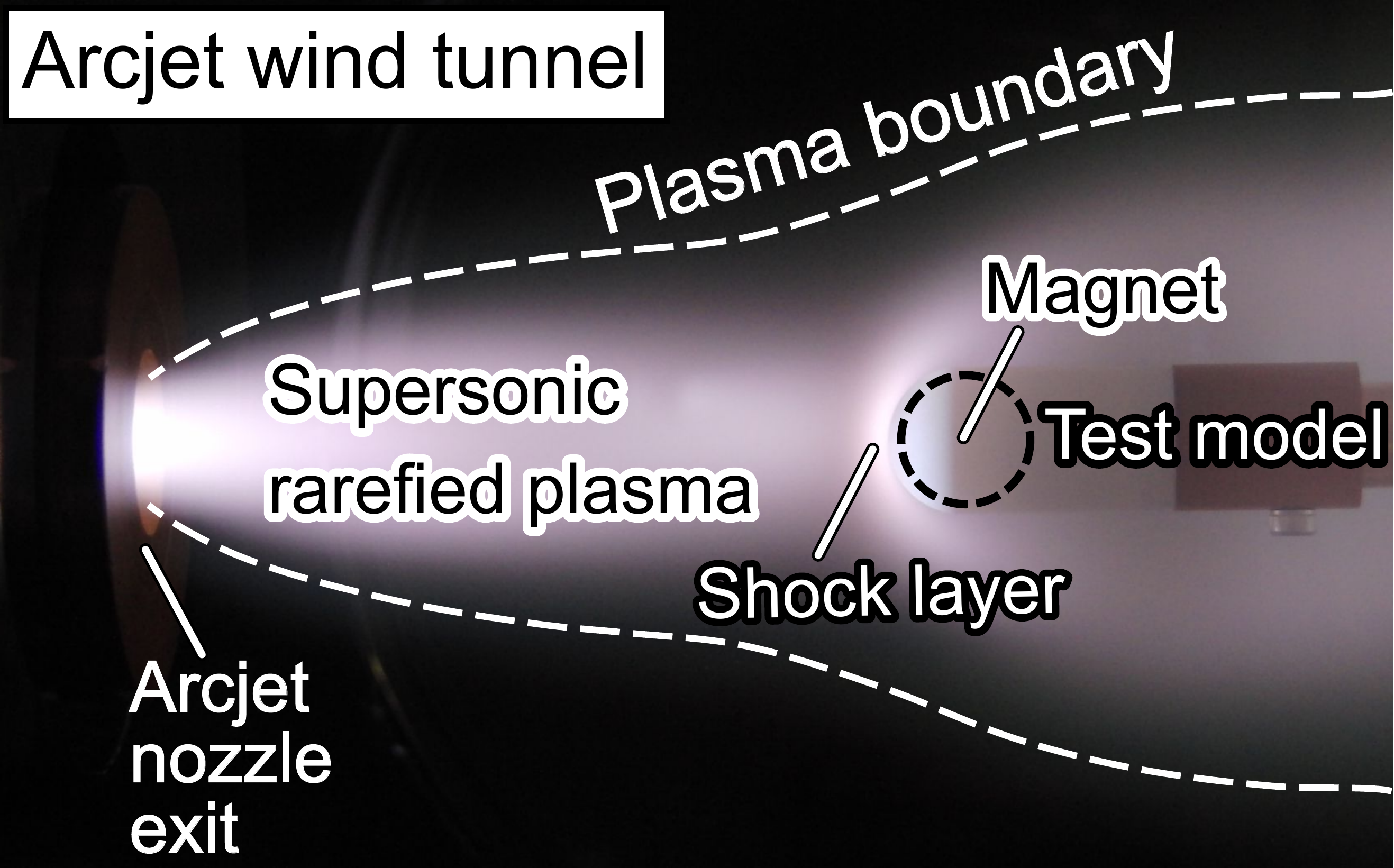}
  \quad
  \includegraphics[height=39mm,clip]{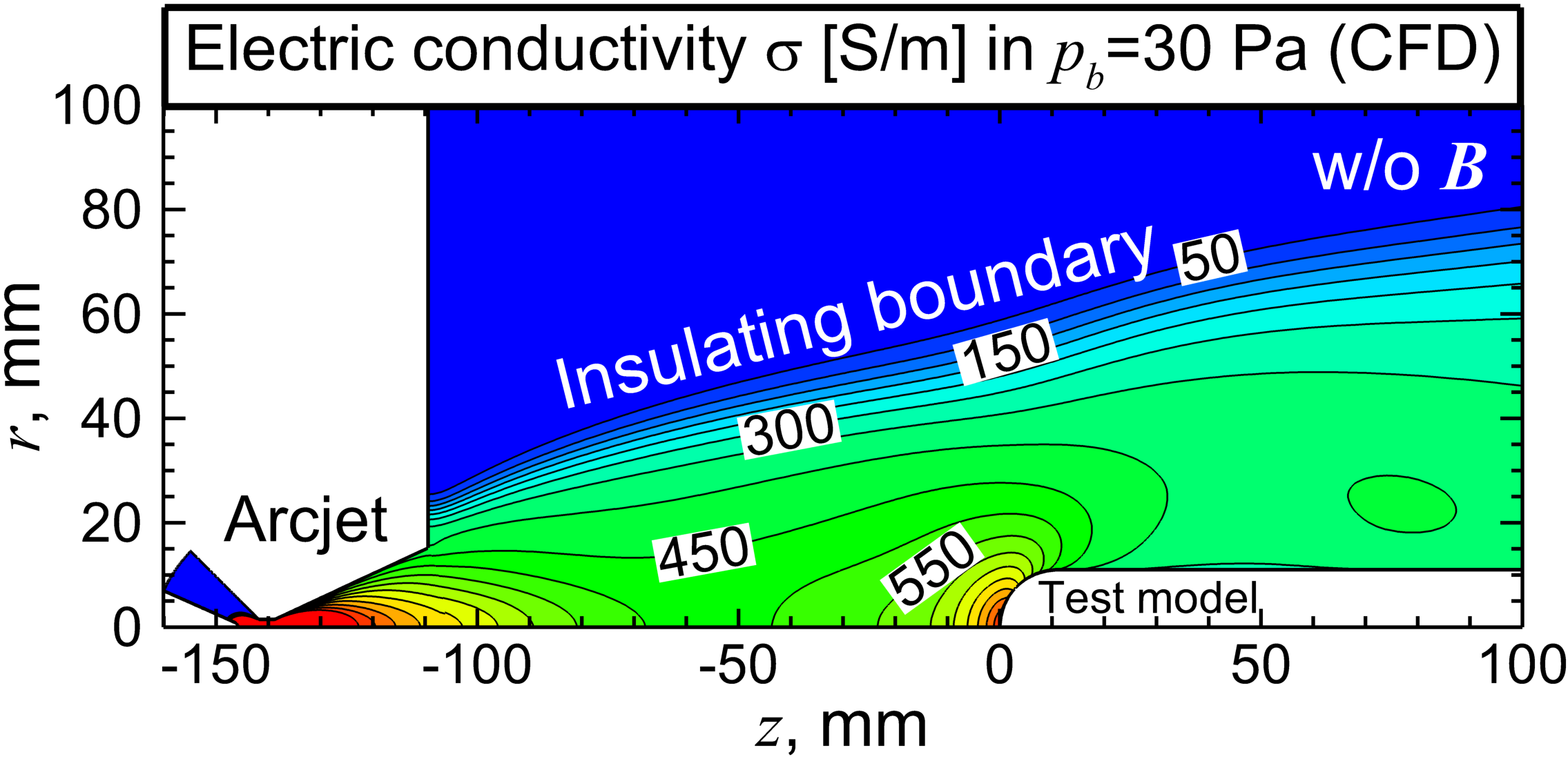}
  \textsf{(a) Experiment} \hspace*{52.5mm} \textsf{(b) CFD simulation} \hspace*{2.5mm}
  \caption{\label{fig:arc} Arcjet wind tunnel experiment (chamber backpressure $p_b\mathord=\SI{30}{Pa}$) of the MHD aerobraking and its CFD simulation (electric conductivity contours in the case without $\bm{B}$).}
  \end{figure}

The influence of the Hall effect on the MHD aerobraking has been discussed mainly in Computational Fluid Dynamics (CFD) simulations~\cite{6, 7} of reentry flows, where only the shock layer in front of the vehicle is ionized. These studies showed that the generation of $\bm{E}$ requires preventing current to the vehicle surface; therefore, the surface must be insulating. On the other hand, to investigate the MHD aerobraking in rarefied flows, we have performed arcjet wind tunnel experiments with argon (figure~\ref{fig:arc}) and demonstrated the MHD shock layer enlargement using absorption spectroscopy~\cite{8, 9} and the drag measurement using a pendulum~\cite{10, 11}. In addition, the detailed CFD simulation~\cite{12} (figure~\ref{fig:arc}) of this experiment predicted that an insulating boundary in the flow, which prevents current dissipation into the flow field, is necessary to generate $\bm{E}$: the plume boundary serves as this insulating boundary in the arcjet wind tunnel, and the ionization front in the shock layer serves as this boundary in the actual reentry flow. In other words, to activate the MHD aerobraking in rarefied flows with the strong Hall effect, the plasma must be sandwiched between the insulating boundaries of the vehicle surface and the plasma boundary in the flow.

\section{Experimental method and setup}
\label{sec:exp_method}

\subsection{Arcjet wind tunnel and magnetized model}
\label{sub_sec:arc_and_model}

We used a popular non-transferred argon arcjet, consisting of the axisymmetric components of a cathode, anode, constrictor, and conical nozzle with a half-angle \ang{25} connected to a vacuum chamber, as the plasma generator for the arcjet wind tunnel (figure~\ref{fig:arc}). The operating argon mass flow rate, discharge current, and discharge voltage of the arcjet are \SI{8}{slm}, \SI{80}{A}, and \SI{15}{V}, respectively. The details of the arcjet can be found in~\cite{13}. The evacuation system with a mechanical booster pump and a rotary pump evacuates the vacuum chamber at \SI{2,100}{m^3/h} and keeps the chamber back pressure $p_b$ at \SI{30}{Pa} under the above operating condition.

\begin{figure}[htbp]
  \centering
  \includegraphics[height=60mm,clip]{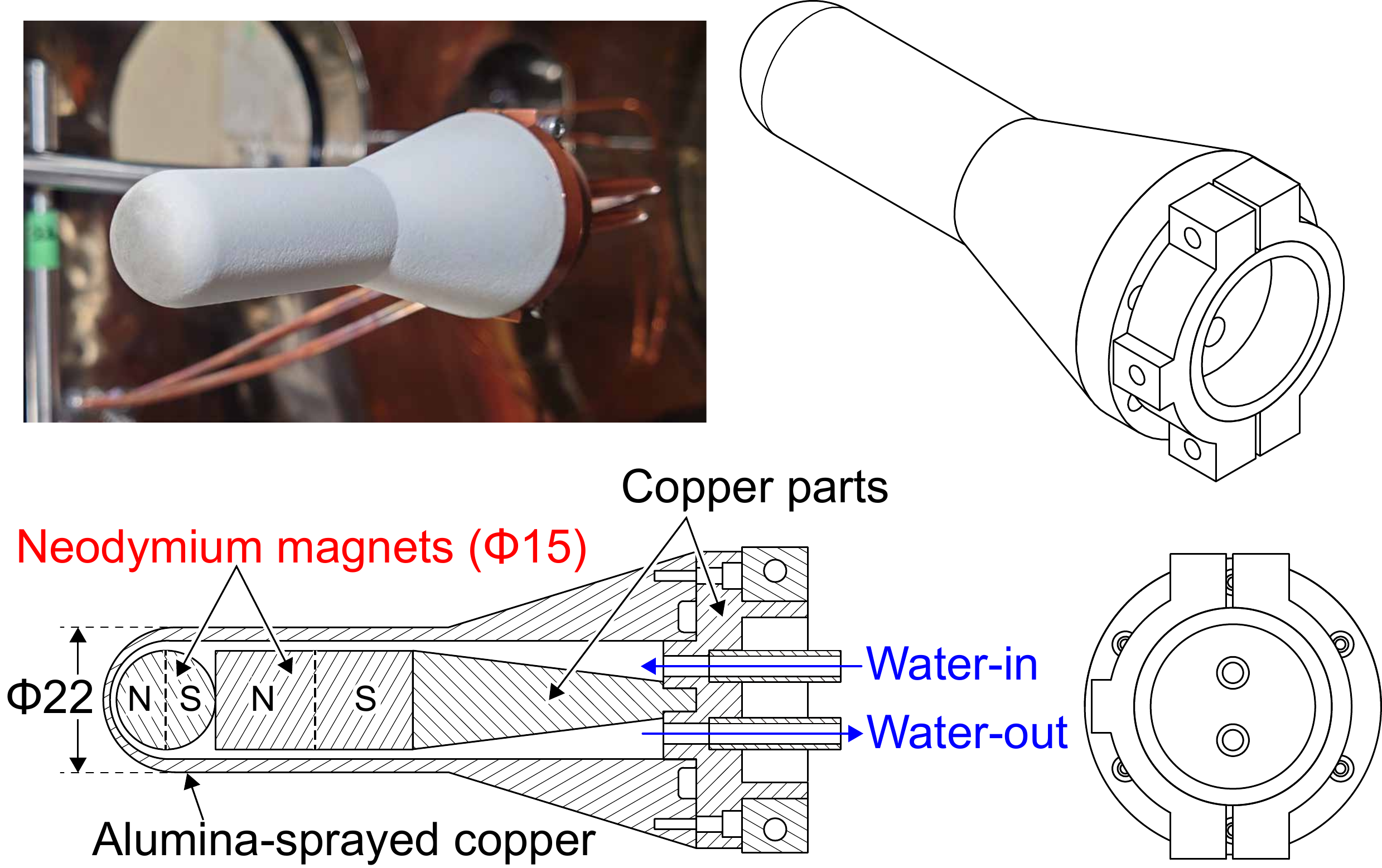}
  \caption{\label{fig:model} Water-cooled magnetized model for the LTS measurement.}
\end{figure}

Figure~\ref{fig:model} shows the magnetized model developed for the LTS measurement. The cover of the model is made of copper and has been sprayed with alumina to give the insulating property to its surface. The diameter of the model is $D=\SI{22}{mm}$. The model mounts tandem neodymium magnets consisting of a sphere and cylinders with a diameter of \SI{15}{mm}. The magnetic pole is aligned with the axis of the model. The model has a magnetic flux density of \SI{0.35}{T} at its stagnation point. The model's interior was water-cooled to prevent thermal demagnetization of the magnet during the measurement time of up to \SI{5}{min}. We set up the stagnation point of the model at \SI{110}{mm} from the nozzle exit (figure~\ref{fig:LTS_sys}). For comparison, we also conducted the experiment with completely demagnetized magnets (i.e., without $\bm{B}$) instead of the magnets.

\subsection{Laser Thomson scattering system}
\label{sub_sec:LTS_sys}

Thomson scattering is characterized by the following parameter $\alpha$~\cite{14},
\begin{equation}
  \label{eq:alpha}
  \alpha = \frac{1}{k \lambda_D} = \frac{\lambda_i}{4\pi \sin\left(\theta/2\right)}\sqrt{\frac{n_e e^2}{\varepsilon_0 k_B T_e}}
\end{equation}
where $k$ is the magnitude of the difference between the incident probe laser and the Thomson scattering light wave vectors, $\lambda_D$ is the Debye length, $\theta$ is the scattering angle to the incident laser wave vector,  $\lambda_i$ is the incident laser wavelength, $\varepsilon_0$ is the permittivity of vacuum, $k_B$ is the Boltzmann constant, $T_e$ is electron temperature. For $\alpha \mathord\ll 1$, the scattering is in the non-collective regime and its spectral shape reflects the electron velocity distribution; therefore, we can obtain $n_e$ from the spectral signal intensity and $T_e$ from the spectral full width at half maximum (FWHM)~\cite{14}. Our CFD simulation~\cite{13} predicted that the representative $n_e$ and $T_e$ are approximately \SI{5e19}{m^{-3}} and \SI[group-minimum-digits=4]{5000}{K} around the stagnation point of the model in the arcjet plume of figure~\ref{fig:arc}. Because $\alpha$ is estimated to be $\mathord\sim 0.1$ from these values, $\lambda_i\mathord=\SI{532}{nm}$, and $\theta\mathord=\ang{90}$, the LTS spectra from the arcjet plume are in the non-collective regime, and we can obtain $n_e$ and $T_e$ from them.

\begin{figure}[b]
  \centering
  \includegraphics[width=\textwidth,clip]{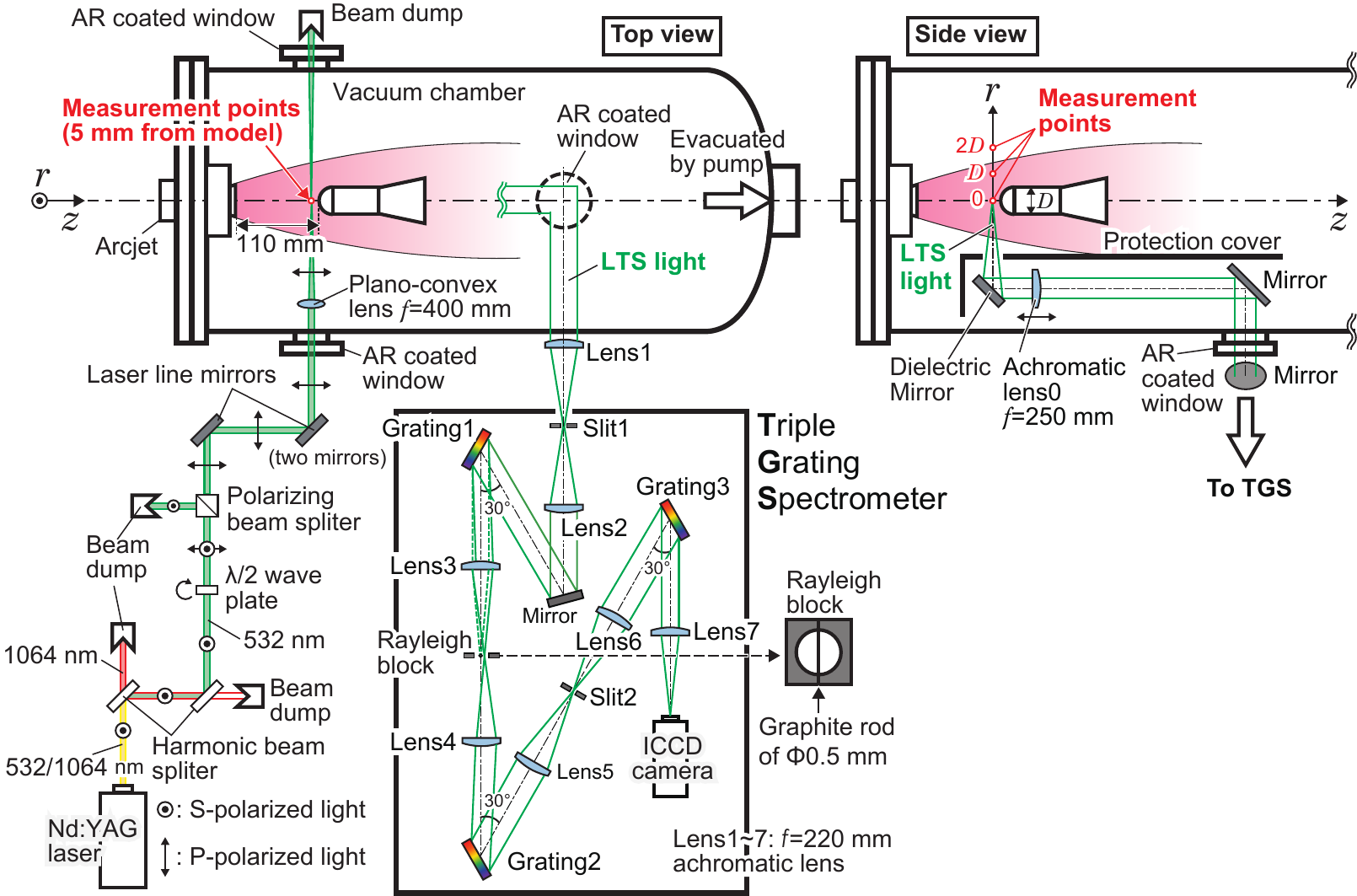}
  \caption{\label{fig:LTS_sys} Laser Thomoson scattering (LTS) measurement system and triple grating spectrometer (TGS).}
\end{figure}

Figure~\ref{fig:LTS_sys} shows our LTS measurement system. We used the second harmonic beam ($\lambda_i\mathord=\SI{532}{nm}$) of the Nd:YAG laser (Continuum, Surelite III with a pulse width of \SI{5}{ns} and repetition rate of \SI{10}{Hz}) as the probe light. Since the irradiated laser contains \SI[group-minimum-digits=5]{1064}{nm} light, only \SI{532}{nm} light is extracted using two harmonic beam splitters. The \SI{532}{nm} light from the laser is S-polarized, and it is rotated using a $\uplambda/2$ waveplate. A polarizing beam splitter separates the rotated light into S and P polarizations. Only the P-polarized light is directed into the vacuum chamber through three laser line mirrors (one vertical and two horizontal reflections) and an AR-coated window because its polarization direction should be perpendicular to the plane containing the probe and scattering light wave vectors to maximize the LTS signal intensity in the radial direction ($r$) including the collection direction with $\theta\mathord=\ang{90}$ of eq.~(\ref{eq:alpha}). The P-polarized probe light is focused onto the measurement point by a plano-convex lens with a focal length of $f\mathord=\SI{400}{mm}$ in the chamber. We adjusted the energy of the probe light by rotating the angle of the waveplate and set it to $E_{i,\mathrm{TS}}\mathord=\SI{150}{mJ}$, at which the plasma heating by the probe light is negligible. The LTS light from the measurement point is collected by a dielectric mirror and achromatic lens (lens0) with $f\mathord=\SI{250}{mm}$, and it is directed to the outside of the chamber through mirrors and an AR-coated window.

The LTS light is focused onto to the entrance slit (slit1, width \SI{0.15}{mm}) of the Triple Grating Spectrometer (TGS) by an achromatic lens (lens1) with $f\mathord=\SI{220}{mm}$. The TGS is a unique handmade spectrometer that can detect the weak LTS light by blocking the stray light and intense Rayleigh scattering light. The details of the TGS can be found in~\cite{15,16,17}. The TGS mainly consists of six achromatic lenses (lens2--7) with $f\mathord=\SI{220}{mm}$, three high modulation holographic reflection gratings (\SI{2,400}{grooves/mm}), a Rayleigh block (width \SI{0.5}{mm}) at the first image plane, and an intermediate slit (slit2, width \SI{0.15}{mm}) at the second image plane. The TGS disperses the light through the first grating and focuses it on the Rayleigh block, which blocks the light at $\lambda\mathord=\SI{532}{nm}$. The light passing through the Rayleigh block is then sent to the second grating, recombining the dispersed spectra into a single beam. The light is then refocused onto the intermediate slit (slit2). The light passing through the slit is dispersed again by the third grating and detected by an intensified charge-coupled device (ICCD) camera (Princeton Instruments, PI-MAX4-1024f with GenIII HBf with quantum efficiency \SI{50}{\%} at \SI{532}{nm}). The gate width of the ICCD camera is set to \SI{7}{ns} and is synchronized with the probe laser with a pulse width of \SI{5}{ns}.

We set the measurement point at $z\mathord=\SI{-5}{mm}$ (axial direction) in front of the stagnation point of the model. This point is inside the shock layer on the stagnation line. We moved the measurement point in the radial direction ($r$) by moving the lens0 (in the vacuum chamber) in the $z$ direction and by moving the probe laser path in the $r$ direction by adjusting the heights of the plano-convex lens and the laser line mirror outside the chamber. We obtained the LTS signals at the stagnation line ($r\mathord=\SI{0}{mm}$), a distance equal to the model diameter ($r\mathord=D$), and a distance double that ($r\mathord=2D$). 

Data accumulation is widely used in LTS measurements to capture weak LTS signals with a good signal-to-noise ratio~\cite{15,16,17}. Because the LTS signal intensity decreases with decreasing electron density in the radial direction, we accumulated the laser-shots (i.e., accumulation count $N_\mathrm{TS}$), irradiated at \SI{10}{Hz}, over $N_\mathrm{TS}\mathord=\SI{100}{shots}$ (\SI{10}{sec}) at the stagnation line ($r\mathord=\SI{0}{mm}$), \SI{300}{shots} (\SI{30}{sec}) at $r\mathord=D$, and \SI{3000}{shots} (\SI{5}{min}) at $r\mathord=2D$.

In our preliminary study~\cite{18}, which measured a single point in front of a non-water-cooled model, we found that subtle distortions of the optical table and optomechanical components due to room temperature variations during the experimental period (sometimes a few days) could lead to misalignment of the optical system, whose path length from the measurement point to the ICCD camera is approximately \SI{5.7}{m}. Since the measured electron density depends on the intensity ratio of Thomson scattering to Rayleigh scattering (discussed in section~\ref{sub_sec:LTS_result}), this misalignment degrades the accuracy of the measured electron density. To overcome this problem, we used an air conditioner to keep the room temperature constant throughout the experimental period. Moreover, to improve the reproducibility of the arcjet flow itself, we disassembled and cleaned the arcjet and polished the cathode tip before the experiment.

\section{Results and discussions}
\label{sec:result}

\subsection{Laser Thomson scattering results}
\label{sub_sec:LTS_result}

\begin{figure}[b]
  \centering
  \includegraphics[width=\textwidth,clip]{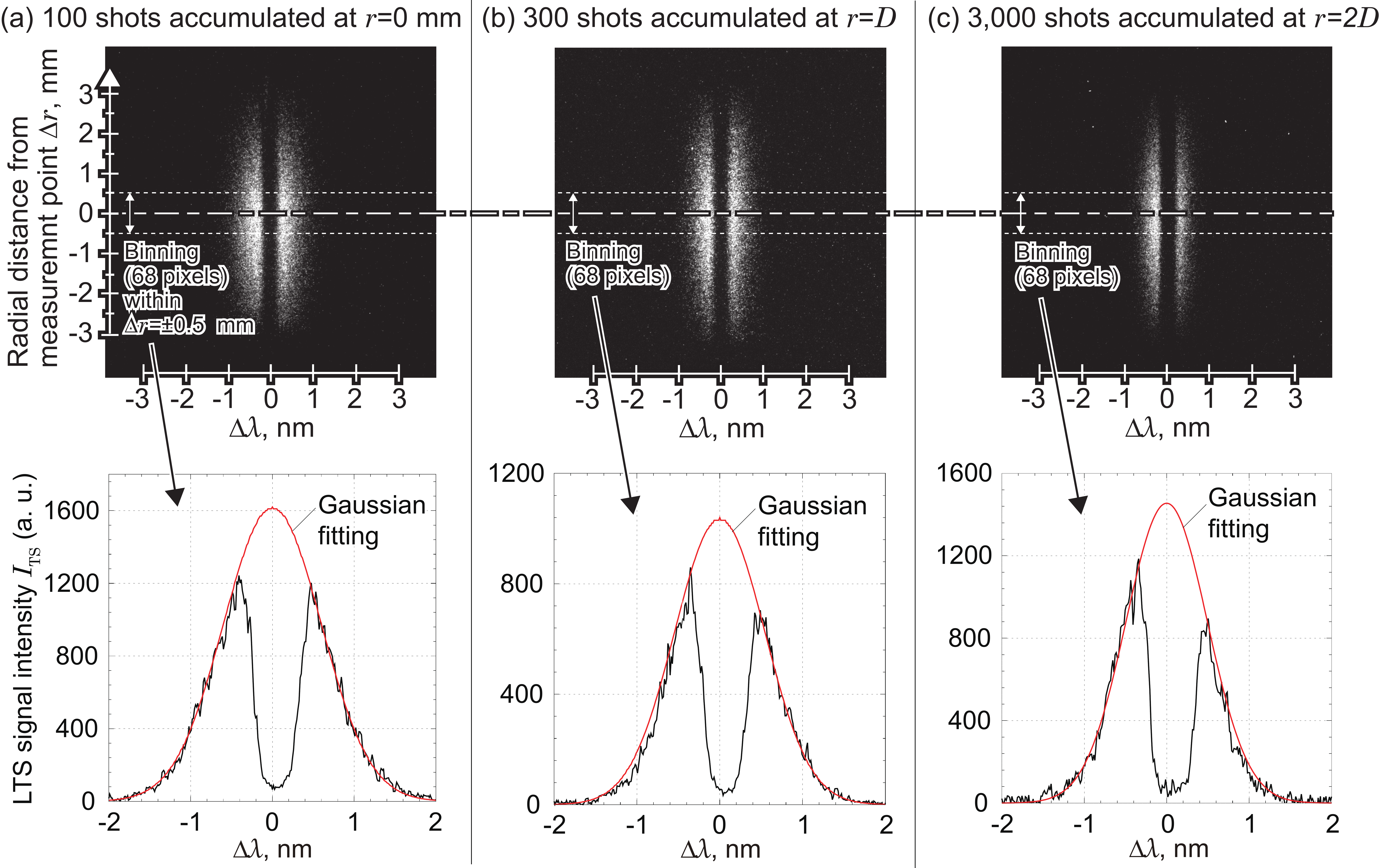}
  \caption{\label{fig:LTS_result} LTS results at $r\mathord=\SI{0}{mm}$, $D$, and $2D$ in the case without $\bm{B}$.}
\end{figure}

Figure~\ref{fig:LTS_result} shows the LTS signal images, whose vertical and horizontal directions are the spatial radial direction and the wavelength sift $\Delta\lambda\left(\mathord=\lambda - \lambda_i\right)$, their binning signal intensities $I_\mathrm{TS}$ within the radial distance from the measurement point $\Delta r\mathord=\pm\SI{0.5}{mm}$ (\SI{68}{pixels}), and their Gaussian fitting curves at $r\mathord=\SI{0}{mm}$ ($N_\mathrm{TS}\mathord=\SI{100}{shots}$), $D$ ($N_\mathrm{TS}\mathord=\SI{300}{shots}$), and $2D$ ($N_\mathrm{TS}\mathord=\SI{3000}{shots}$) in the case without $\bm{B}$. The dark region around $\Delta\lambda\mathord=\SI{0}{nm}$ is the shadow of the Rayleigh block.

Because the non-collective LTS intensity $I_\mathrm{TS}$ is proportional to its dynamic form factor $S_\mathrm{TS}$ expressed as follows~\cite{19},
\begin{equation}
  \label{eq:I_TS}
  I_\mathrm{TS}\left(\Delta\lambda\right) \propto S_\mathrm{TS}\left(\Delta\lambda\right) = \left(\frac{m_e}{2\pi k_B T_e}\right)^{1/2}\left[\frac{c}{2\lambda_i \sin\left(\theta / 2\right)}\right]\exp\left\{-\frac{m_e}{2 k_B T_e}\left[\frac{c\Delta\lambda}{2\lambda_i \sin\left(\theta/2\right)}\right]^2\right\}
\end{equation}
which can be fitted by the Gaussian function $G\left(\Delta\lambda\right)\mathord=A \exp\left(-B \Delta\lambda^2\right)$ with parameter $A$ and $B$, we can obtain $T_e$ from
\begin{equation}
  \label{eq:T_e}
  T_e = \frac{1}{B}\frac{m_e}{2 k_B}\left[\frac{c}{2\lambda_i\sin\left(\theta/2\right)}\right]^2
\end{equation}
where $m_e$ and $c$ are the electron mass and the speed of light. To obtain $n_e$, we calibrate the absolute value of $I_\mathrm{TS}$ using the Rayleigh scattering intensity $I_\mathrm{RS}$ in nitrogen molecule of room temperature~\cite{15,16,17} as follows~\cite{14,19,20},
\begin{equation}
  \label{eq:n_e}
  n_e = \frac{n_0}{I'_\mathrm{RS}}I'_\mathrm{TS}\frac{E_{i,\mathrm{RS}}}{E_{i,\mathrm{TS}}}\frac{\sigma_\mathrm{RS}}{\sigma_\mathrm{TS}}\frac{N_\mathrm{RS}}{N_\mathrm{RS}}
\end{equation}
where $I'_\mathrm{TS}$ and $I'_\mathrm{RS}$ are the total Thomson and Rayleigh scattering intensities which are summed over the wavelength direction as $I'_\mathrm{TS}\mathord=\sum_{\mathrm{ovearall} \, \Delta\lambda} I_\mathrm{TS}$  and $I'_\mathrm{RS}\mathord=\sum_{\mathrm{ovearall} \, \Delta\lambda} I_\mathrm{RS}$, $E_{i,\mathrm{TS}}$, $E_{i,\mathrm{RS}}$, $N_\mathrm{TS}$, and $N_\mathrm{RS}$ are the probe laser energies and accumulation count for the Thomson and Rayleigh scattering measurements, the ratio of the Thomson to the Rayleigh scattering cross-sections of nitrogen molecule $\sigma_\mathrm{TS}/\sigma_\mathrm{RS}$ is 131 with $\lambda_i\mathord=\SI{532}{nm}$, and $n_0$ is nitrogen molecule number densities. We measured the Rayleigh scattering light at the points consistent with the LTS measurements using the LTS system (figure~\ref{fig:LTS_sys}), which is without the Rayleigh block in the TGS, in the vacuum chamber filled with nitrogen molecule at room temperature and 0 to \SI{300}{Torr}. In the Rayleigh scattering measurement, we set $E_{i,\mathrm{RS}}$ and $N_\mathrm{RS}$ to \SI{8}{mJ} and \SI{5}{shots}, and measured $n_0$ by a capacitance manometer (MegaTorr CDLD-31S06J with full-scale \SI{1000}{Torr} and \SI{0.5}{\%} reading accuracy) on the chamber wall. The ratios $I'_\mathrm{RS}/n_0$ obtained from three measurements at each measurement point are \num{6.61e-20}, \num{5.32e-20}, and \SI{5.72e-20}{{(a.u.)}\cdot m^3} at $r\mathord=\SI{0}{mm}$, $D$, and $2D$, respectively. We used these values in eq.~(\ref{eq:n_e}).

\subsection{Measured electron temperature and density, and comparison with CFD}
\label{sub_sec:comp_exp_vs_CFD}

\begin{figure}[b]
  \centering
  \includegraphics[width=\textwidth,clip]{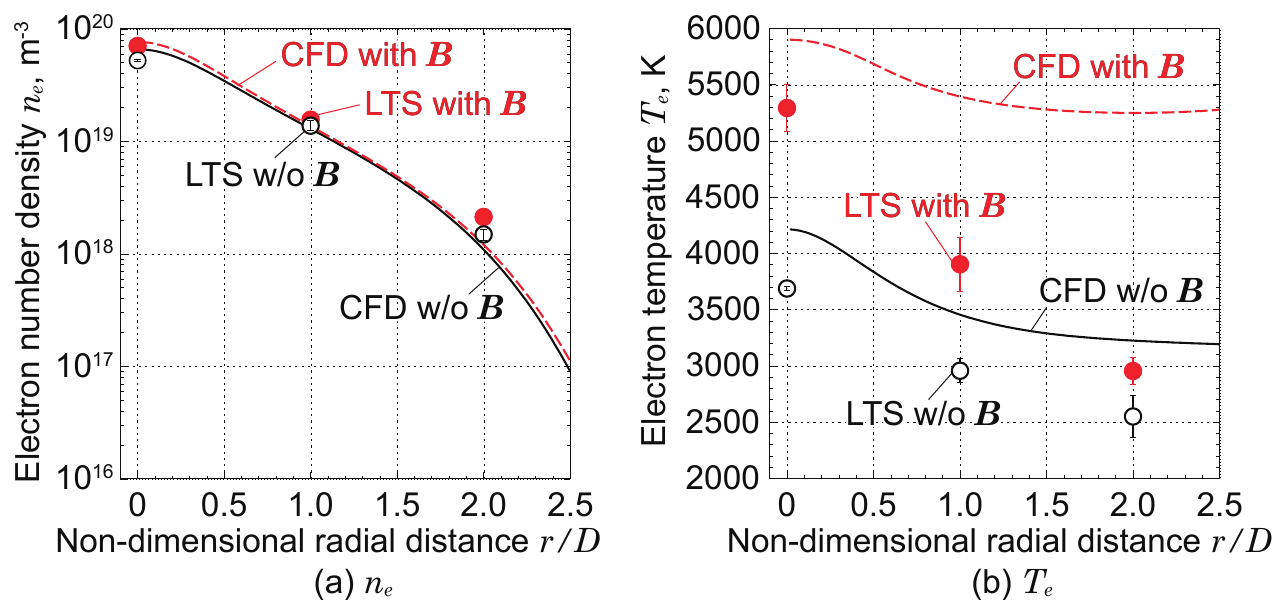}
  \caption{\label{fig:comp_exp_vs_cfd} Radial distributions of (a) $n_e$ and (b) $T_e$ at $z=\SI{-5}{mm}$ in front of the model: LTS measurements and CFD predictions in the cases with and without $\bm{B}$.}
\end{figure}

Figure~\ref{fig:comp_exp_vs_cfd} shows the electron density and temperature obtained from the LTS measurements and the CFD predictions~\cite{12} recomputed for the present experimental condition. We performed three measurements at each measurement point to evaluate the error bars.

The electron density showed excellent agreement between the measurements and CFD predictions in both cases with and without $\bm{B}$. It was also found that $\bm{B}$ had little effect on the distribution of $n_e$. In the case without $\bm{B}$, although the measured $T_e$ was about \SI{500}{K} lower than the CFD prediction, its decreasing trend in the radial direction was in good agreement with that of the CFD. These agreements suggests that the CFD faithfully reproduces the flow field near the model in the experiment.

As mentioned in section~\ref{sec:intro}, when $\bm{B}$ is applied, the CFD predicts that $T_e$ increases due to the Joule heating caused by the Hall electric field $\bm{E}$, and the increase is about \qtyrange[range-phrase= --, range-units= single]{1500}{2000}{K} in the present conditions. The measurements also show a similar increase of $T_e$ by applying $\bm{B}$, but the increased range is about \qtyrange[range-phrase= --, range-units= single]{500}{1500}{K}, which is much smaller than that of the CFD. This discrepancy may be due to the location of an insulating boundary far from the model in the experiment: The far-field insulating boundary allows current dissipation into the plasma flow field, weakening the Hall electric field and the resulting Joule heating. Moreover, this indicates that the CFD continuum approximation might not be acceptable near the plume boundary, and any particle simulation will be necessary to reproduce the present experiment correctly.

\section{Conclusion}
\label{sec:conclusion}

To elucidate the role of the Hall effect in the MHD aerobraking in rarefied flows, the electron temperature and density around a magnetized model were measured in a rarefied argon arcjet wind tunnel using the laser Thomson scattering method. The plasma in this study is in the non-collective regime, and we measured the Thomson scattering light using a handmade triple grating spectrometer with a Rayleigh block and obtained the radial distribution of electron temperature and density around the magnetized body with high accuracy. We also developed a water-cooled magnetized model to prevent thermal demagnetization during the measurement for up to five minutes, in which we perform several thousands of data accumulations to capture the weak LTS signals with a good signal-to-noise ratio.

The measured electron density distribution is in excellent agreement with the CFD prediction. It was also found that the magnetic field had little effect on the electron density distribution. In the case without the magnetic field, although the measured electron temperature was about \SI{500}{K} lower than the CFD prediction, its decreasing trend in the radial direction was in good agreement with that of the CFD. These agreements suggests that the CFD faithfully reproduces the flow field near the model in the experiment.

However, in the case with the magnetic field, the measured electron temperature increase was about \SI{1000}{K} less than that of the CFD. This discrepancy may be due to the location of an insulating boundary far from the model in the experiment: The far-field insulating boundary allows current dissipation into the plasma flow field, weakening the Hall electric field and the resulting Joule heating. Moreover, this indicates that the CFD continuum approximation might not be acceptable near the plume boundary, and any particle simulation will be necessary to reproduce the present experiment correctly.

\acknowledgments

This work was supported by JSPS KAKENHI Grant Numbers JP15H04200, JP16K14505, JP19H02348, JP19K22018, JP21KK0078, JP22K04534.

\end{document}